\documentclass{aastex6}
\usepackage{multirow}
\begin{document}

\title{On the Energy Sources of the Most Luminous Supernova ASASSN-15lh}
\author{Long Li\altaffilmark{1,2}, Zi-Gao Dai\altaffilmark{1,2}, Shan-Qin Wang\altaffilmark{3}, and Shu-Qing Zhong\altaffilmark{1,2}}

\begin{abstract}
In this paper, we investigate the energy-source models for the most luminous supernova ASASSN-15lh. We revisit the ejecta-circumstellar medium (CSM) interaction (CSI) model and the CSI plus magnetar spin-down with full gamma-ray/X-ray trapping which were adopted by \cite{Chatzopoulos16} and find that the two models cannot fit the bolometric LC of ASASSN-15lh. Therefore, we consider a CSI plus magnetar model with the gamma-rays/X-rays leakage effect to eliminate the late-time excess of the theoretical LC. We find that this revised model can reproduce the bolometric LC of ASASSN-15lh. Moreover, we construct a new hybrid model (i.e., the CSI plus fallback model), and find that it can also reproduce the bolometric LC of ASASSN-15lh. Assuming that the conversion efficiency ($\eta$) of fallback accretion to the outflow is typically $\sim10^{-3}$, we derive that the total mass accreted is $\sim3.9~M_\odot$. The inferred CSM mass in the two models is rather large, indicating that the progenitor could have experienced an eruption of hydrogen-poor materials followed by an energetic core-collapse explosion leaving behind a magnetar or a black hole.
\end{abstract}


\affil{\altaffilmark{1}School of Astronomy and Space Science, Nanjing
University, Nanjing 210093, China; dzg@nju.edu.cn}
\affil{\altaffilmark{2}Key Laboratory of Modern Astronomy and Astrophysics
(Nanjing University), Ministry of Education, China}
\affil{\altaffilmark{3}Guangxi Key Laboratory for Relativistic Astrophysics,
School of Physical Science and Technology, Guangxi University, Nanning 530004, China; shanqinwang@gxu.edu.cn}

\section{Introduction}
\label{sec:intro}

In the past two decades, more than 100 super-luminous supernovae (SLSNe) have been found \citep{GalYam12,GalYam19,Inserra19} by several sky-survey projects for optical transients (see, e.g., \citealt{Chomiuk11,Quimby11,Nicholl14,Quimby14,DeCia18,Lunnan18,Angus19}). Just like normal supernovae (SNe), SLSNe can be classified as types I and II, depending on whether the spectra  contains hydrogen absorption lines or not. In addition to their luminosity, the main difference between SLSNe and normal SNe is their energy sources. Most normal SNe can be explained by the $^{56}$Ni cascade decay model that doesn't apply to almost all SLSNe. The energy sources of SLSNe are still elusive. The pair instability SN (PISN) model \citep{Barkat67,Rakavy67,Heger02,Heger03} was suggested to account for some SLSNe, while the magnetar spin-down model \citep{Kasen10,Inserra13,Nicholl13,Nicholl14,Wang15,Wanglj16}, the ejecta-circumstellar medium (CSM) interaction (CSI) model \citep{Chevalier82,Chevalier94,Chevalier11,Chatzopoulos12,Liu18}, and the fallback model \citep{Dexter13} were used to explain most of SLSNe (see \citealt{Moriya18,Wang19} and references therein).

To date, the most luminous SLSN might be ASASSN-15lh, which was an extremely luminous optical-UV transient discovered by All-Sky Automated Survey for SuperNovae (ASAS-SN, \citealt{Shappee14})
on June 14, 2015 \citep{Dong16}. At early times ($t<30$ days, rest frame adopted throughout), there are only $V$ band data observed by ASAS-SN. The multi-band follow-up observations were performed by the $V$ and $B$ filters of the Las Cumbres Observatory Global Telescope Network (LCOGT; \citealt{Brown13}) and $V$, $B$, $U$, $UVW1$, $UVM2$, and $UVW2$ filters of the UltraViolet and Optical Telescope (UVOT) on board \emph{the Neil Gehrels Swift Observatory} (\emph{Swift}, \citealt{Gehrels04,Roming05}). The UV observations performed by UVOT last from 30 to 450 days except for the sun constraint break ($230-240$ days). The UVOT $V$- and $B$-band observations were sometimes interrupted, but the LCOGT $V$- and $B$-band observations were always used when needed.\footnote{\cite{GodoyRivera17} translated the V- and B-band magnitudes of LCOGT to the \emph{Swift} magnitude system.}

Based on the blackbody assumption, \cite{Dong16} and \cite{GodoyRivera17} used the multi-band LCs to fit the early-time and the whole bolometric LCs, respectively.
For the very early epoch ($t<30$ days), \cite{GodoyRivera17} adopted two different evolution modes of the blackbody temperature, i.e., a linearly increasing temperature in a logarithmic scale and a constant temperature, to obtain the corresponding bolometric LCs. At a redshift of $z=0.2326$, ASASSN-15lh reached a peak bolometric luminosity of $(2.2\pm0.2)\times 10^{45}\rm\ erg\ s^{-1}$, more than twice as luminous as any previously known SNe. After the main peak lasted for $\sim90$ days, there was a rebrightening of \emph{Swift} UV bands, and the bolometric LC showed a $\sim120$ days plateau, and then faded again \citep{GodoyRivera17}. The total radiation energy of ASASSN-15lh is $\sim1.7-1.9\times10^{52}$ erg over the $\sim450$ days since the first detection \citep{GodoyRivera17}. Besides, at the location of ASASSN-15lh, a persistent X-ray emission whose luminosity is $\sim10^{41}-10^{42}\rm\ erg\ s^{-1}$ was observed by the Chandra X-ray observatory \citep{Margutti17}.

The nature of ASASSN-15lh is still in debate. \cite{Dong16} classified ASASSN-15lh as a hydrogen-poor (type I) SLSN;
after analyzing the entire evolution of photospheric radius as well as the radiated energy and estimating the event rate,
\cite{GodoyRivera17} thought ASASSN-15lh is more similar to a H-poor SLSN rather than a TDE.
On the other hand, \cite{Leloudas16} and \cite{Kruhler18} claimed that it is a tidal disruption event (TDE).

The extremely high peak luminosity, long duration, and exotic bolometric LC challenge all existing energy-source models for SLSNe. \cite{Dong16} estimated that at least $30~M_\odot$ of $^{56}$Ni is required to produce the observed peak luminosity of ASASSN-15lh if the LC was powered by $^{56}$Ni cascade decay, while \cite{Kozyreva16} showed that $1500~M_\odot$ of $^{56}$Ni is needed to power the bolometric LC based on numerical simulation. Some other authors \citep{Metzger15,Dai16,Bersten16,Sukhbold16} suggested that a magnetar with extremely rapid rotation can drive the early-time extremely luminous bolometric LC of ASASSN-15lh. \cite{Chatzopoulos16} showed that a CSI model with ejecta mass of $\sim35~M_\odot$ and CSM mass of $\sim20~M_\odot$ could reproduce the first $\sim220$ days of the bolometric LC, i.e., the main peak and the plateau.
However, the entire bolometric LC of ASASSN-15lh spanning $\sim450$ days has not been modeled by the models mentioned above.
Recently, \cite{Mummery20} fitted the multi-band light curves of ASASSN-15lh using the TDE model involving a super-massive maximally
rotating black hole whose mass is $\sim 10^9$ M$_\odot$.

In this paper, we investigate several possible energy sources of ASASSN-15lh.
In section \ref{sec:modeling}, three models are used to reproduce the bolometric LC of ASASSN-15lh.
Our discussion and conclusions can be found in Sections \ref{sec:dis} and \ref{sec:con}, respectively.

\section{Modeling the Bolometric LC of ASASSN-15lh}
\label{sec:modeling}

In this section, we use three energy-source models (the CSI model, the CSI plus magnetar model, and the CSI plus fallback model) to fit
the bolometric LC of ASASSN-15lh which is taken from \cite{GodoyRivera17}. It should be noted that the bolometric LC at $t<30$ days is constructed
by assuming a logarithmic linearly increasing temperature \cite{GodoyRivera17}. For each model, both the wind-like CSM ($s=2$) and dense-shell
CSM ($s=0$) are taken into account. The value of the optical opacity $\kappa$ of the hydrogen-poor ejecta and the CSM is fixed to be
$0.2 \rm\ cm^2\ g^{-1}$ throughout this paper, the conversion efficiency of the kinetic energy to radiation is assumed to be 100\%
\citep{Chatzopoulos13}.

We develop our own Python-based semi-analytic models and use them to fit the bolometric LC of ASASSN-15lh. Bayesian analysis is adopted
to determine the best fitting parameters. We use the \texttt{emcee} python package \citep{emcee} based on Markov Chain Monte Carlo (MCMC) by
performing a maximum likelihood $\chi^2$ fit, and provides the posterior probability distributions for the free parameters in these models.
The free parameters and priors in our models can be seen in Table \ref{tab:prior}. We ran the MCMC with 20 walkers for running 100,000 steps.
Once the MCMC is done, the best fit values and the uncertainties are computed as the 16th, 50th, and 84th percentiles of the posterior samples
along each dimension, i.e., the uncertainties are measured as $1\sigma$ confidence ranges.

\subsection{The CSI Model}
\label{subsec:csi}

\cite{Chatzopoulos16} pointed out that the CSI model can explain the first $\sim220$ days bolometric LC of ASASSN-15lh. To verify whether the
model can explain the $\sim450$ days bolometric LC, we use the same model the whole LC. The semi-analytical CSI model adopted here was
discussed by \cite{Chatzopoulos12}, \cite{Chatzopoulos13} and \cite{Wanglj19}. Since the ASASSN-15lh is a type I SN, $n=7$ and $\delta=0$ are
adopted for the ejecta outer and inner density profiles.

\begin{figure}[tbph]
\begin{center}
\includegraphics[width=0.45\textwidth,angle=0]{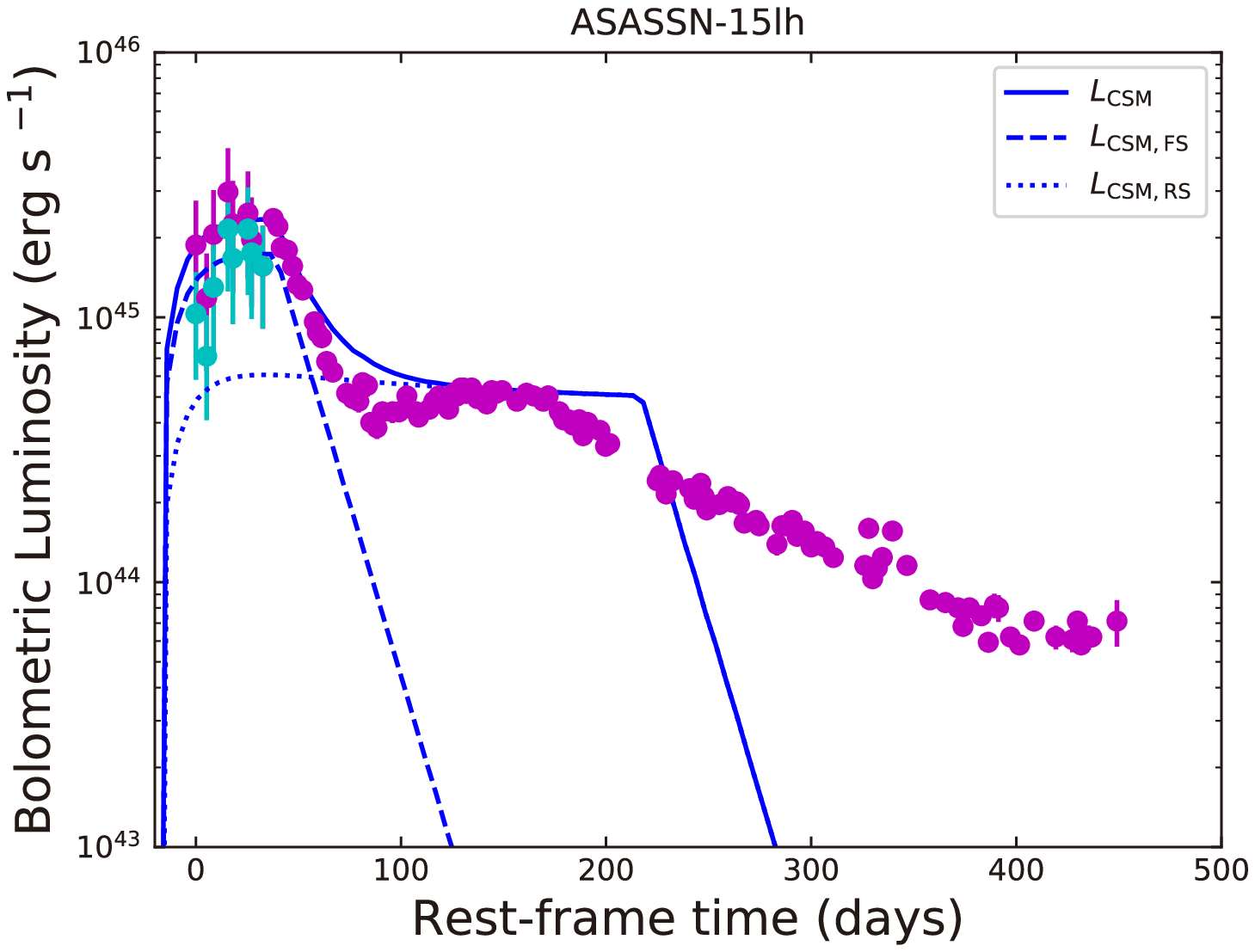}
\includegraphics[width=0.45\textwidth,angle=0]{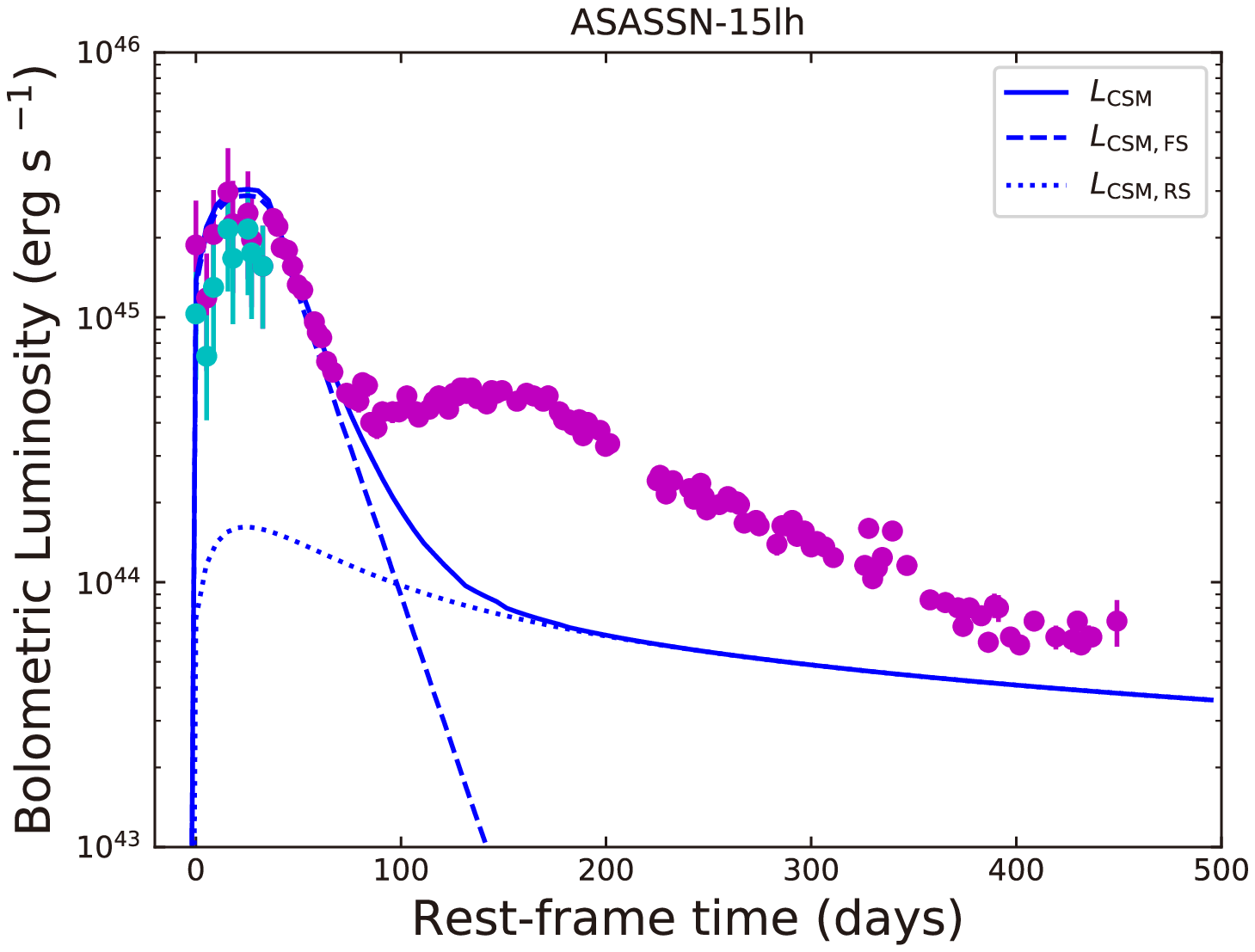}
\end{center}
\caption{The bolometric LCs of ASASSN-15lh reproduced by the CSI model: $s = 0$ (the left panel) and $s = 2$ (the right panel). The total luminosity ($L_{\rm tot}$) are shown by solid blue curves and the contributions from the forward shock ($L_{\rm CSI,FS}$) and the reverse shock ($L_{\rm CSI,RS}$) are shown by dashed and dotted blue curves, respectively. Data are taken from \cite{GodoyRivera17}, where the first eight data points was derived based on the assumptions of the rising-temperature feature (magenta) and constant temperature feature (cyan), respectively.}
\label{fig:CSI}
\end{figure}

Finally, the CSI model has 7 free parameters: $M_{\rm{ej}}$, $v_{\rm{SN}}$, $M_{\rm{CSM}}$, $\rho_{\rm{CSM,in}}$, $R_{\rm CSM,in}$, $x_0$,
and the moment of explosion $t_{\rm{expl}}$. The theoretical bolometric LCs are shown in Figure \ref{fig:CSI}. It can be found that this model
cannot reproduce the entire luminosity evolution: in the case of wind-like CSM, the CSI model can only reproduce the main peak; in the case
of dense-shell CSM, the CSI model cannot reproduce the late-time decline of bolometric LC of ASASSN-15lh.

\subsection{The CSI Plus Magnetar Model}
\label{subsec:mag+csi}

\cite{Chatzopoulos16} also adopted the CSI plus magnetar model to model the first $\sim220$ days LC of ASASSN-15lh.
Here we use the same model to model the whole bolometric LC.

For the semi-analytical CSI plus magnetar model, there are two cases for the output luminosity: a homogeneously expanding photosphere
and a fixed photosphere. The former is applied to some centrally located energy sources (e.g., the $^{56}$Ni cascade decay, the magnetar
spin-down radiation and the fallback accretion outflow) heating the expansive SN ejecta, while the latter is mainly applied to the CSI model,
in which the nearly stationary CSM relative to the ejecta is heated.

We next consider a CSI plus magnetar model to fit the bolometric LC of ASASSN-15lh. We divide the radiative process into two phases:
the early-time fixed-photosphere phase before the ejecta sweeps up the CSM and the late-time homogeneously expanding-photosphere phase
after the ejecta sweeps up the CSM, respectively. We suppose that the CSI dominates the early peak of the bolometric LC before 90 days,
and the late-time plateau and subsequent phase after ejecta sweeps up the optically thick CSM were mainly powered by the magnetar spin-down.
The total ejecta mass at the late epoch becomes $M_{\rm ej}+M_{\rm CSM,th}$. Based on these assumptions, the CSI plus magnetar model
we adopt can be expressed by
\begin{equation}
  L(t)=\frac{1}{t_0} e^{-\frac{t}{t_0}} \int_{t_{\rm expl}}^{t+t_{\rm expl}} e^{\frac{t'}{t_0}} \left[L_{\rm inp,FS}(t')+L_{\rm inp,RS}(t')\right] d t' + \frac{2}{t_d} e^{\frac{t^2}{t_d^2}} \int_{t_{\rm expl}}^{t+t_{\rm expl}} e^{\frac{t'^2}{t_d^2}} L_{\rm inp,mag}(t') d t',
\end{equation}
where $L_{\rm inp,FS}(t)$ and $L_{\rm inp,RS}(t)$ are the input luminosities from
the forward shock and reverse shock, respectively \citep{Chatzopoulos12};
$L_{\rm inp,mag}(t)$ is the input luminosity from the magnetar spin-down \citep{Kasen10},
$t_0$ and $t_d=[\frac{2 \kappa (M_{\rm ej}+M_{\rm CSM,th})}{\beta c v_{\rm SN}}]^{1/2}$ are
the effective LC timescales in a fixed photosphere and an expanding
photosphere, respectively \citep{Chatzopoulos12}.

\begin{figure}[tbph]
\begin{center}
\includegraphics[width=0.45\textwidth,angle=0]{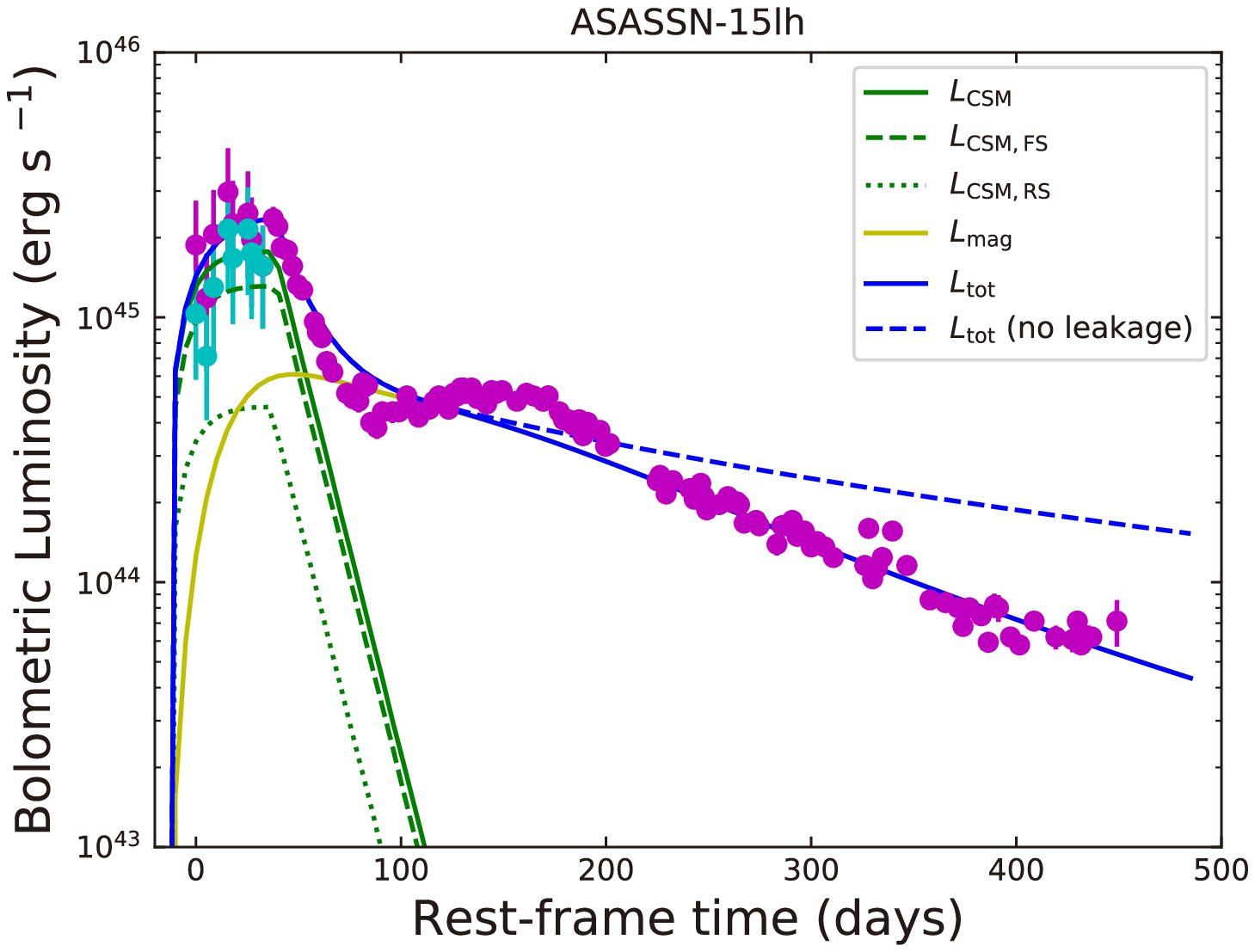}
\includegraphics[width=0.45\textwidth,angle=0]{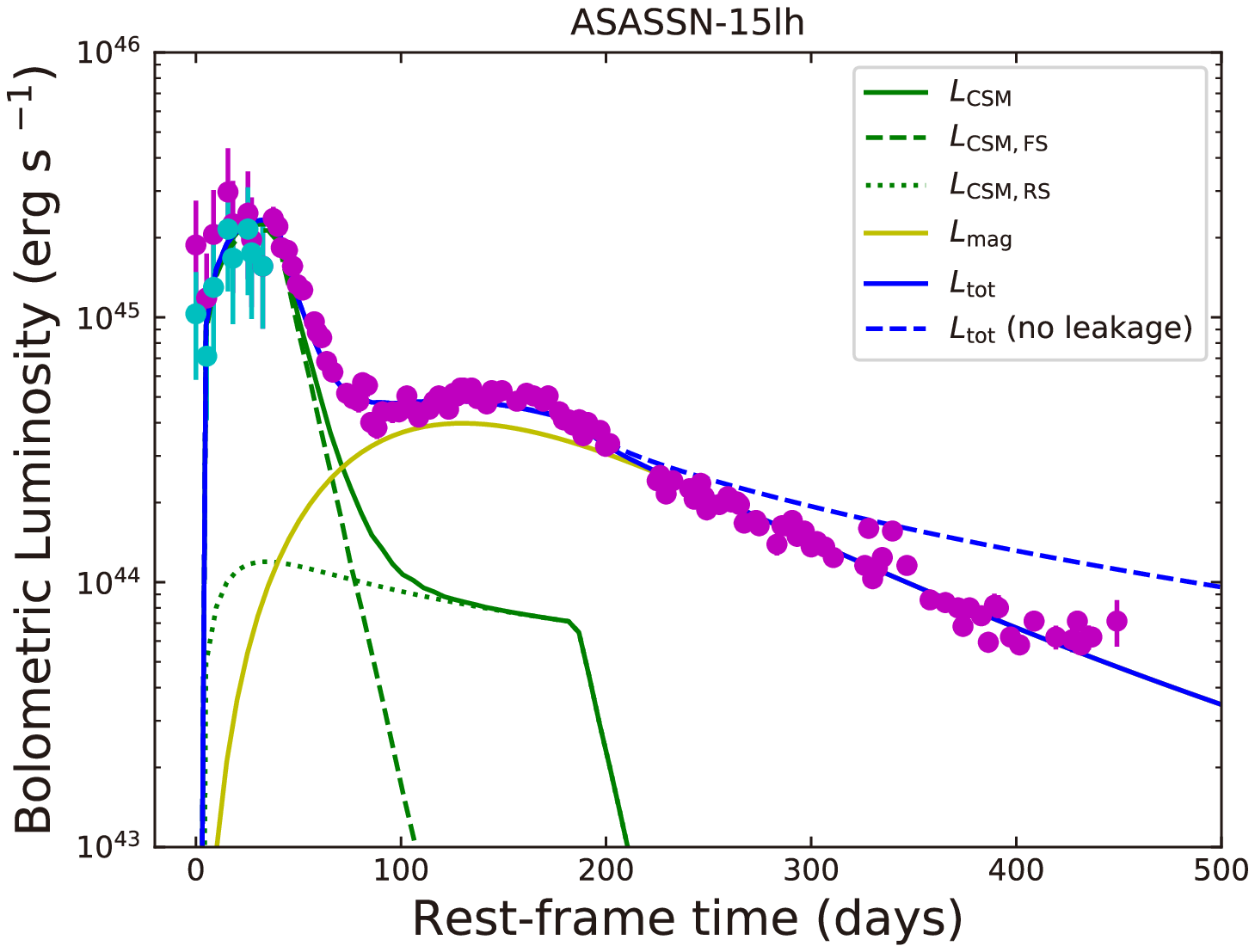}
\end{center}
\caption{The bolometric LCs of ASASSN-15lh reproduced by the CSI plus magnetar model: $s = 0$ (the left panel) and $s = 2$ (the right panel). The total luminosity ($L_{\rm tot}$) are shown by solid blue curves and the contributions from the forward shock ($L_{\rm CSI,FS}$), the reverse shock ($L_{\rm CSI,RS}$), and the magnetar spin-down ($L_{\rm mag}$) are shown in dashed green, dotted green, and solid yellow curves, respectively. The dashed blue curves represent the total luminosity that the leakage effect is not taken into account. Data are taken from \cite{GodoyRivera17}.}
\label{fig:mag+CSI}
\end{figure}

The CSI plus magnetar model has 9 free parameters: $M_{\rm{ej}}$, $v_{\rm{SN}}$, $P_0$, $B_p$,
$M_{\rm{CSM}}$, $\rho_{\rm{CSM,in}}$, $R_{\rm CSM,in}$, $x_0$, and $t_{\rm{expl}}$.
The LC produced by the CSI plus magnetar model are shown in dashed blue lines in Figure \ref{fig:mag+CSI}.
We find that the CSI+magnetar model used by \cite{Chatzopoulos16} cannot reproduce the entire luminosity evolution
since the late-time theoretical LCs produced by both the shell-CSI and the wind-CSI are brighter than the
observations.

To eliminate the late-time excess, we employ a CSI plus magnetar model by considering the
leakage effect of gamma-ray/X-rays from the magnetar \citep{Wang15} which can be described as
\begin{equation}
  L(t)=\frac{1}{t_0} e^{-\frac{t}{t_0}} \int_{t_{\rm expl}}^{t+t_{\rm expl}} e^{\frac{t'}{t_0}} \left[L_{\rm inp,FS}(t')+L_{\rm inp,RS}(t')\right] d t' + \frac{2}{t_d} e^{\frac{t^2}{t_d^2}} \int_{t_{\rm expl}}^{t+t_{\rm expl}} e^{\frac{t'^2}{t_d^2}} L_{\rm inp,mag}(t') \left(1-e^{-\tau_{\rm \gamma,mag}}\right) d t',
\end{equation}
where $e^{-\tau_{\rm \gamma,mag}}$ and $1-e^{-\tau_{\rm \gamma,mag}}$ are the leaking factor and the trapping factor
which represent the gamma-ray/X-ray leakage and trap from the magnetar, respectively; $\tau_{\rm \gamma,mag}$
is the optical depth of the ejecta to gamma-ray/X-ray emissions which can be written as
$\tau_{\rm \gamma,mag} = \frac{3 \kappa_{\rm \gamma,mag} (M_{\rm ej}+M_{\rm CSM})}{4 \pi v_{SN}^2 t^2}$,
$\kappa_{\rm \gamma,mag}$ is the opacity of the gamma-ray/X-ray generated by magnetar spinning-down.

\begin{figure}[tbph]
\begin{center}
\includegraphics[width=1.0\textwidth,angle=0]{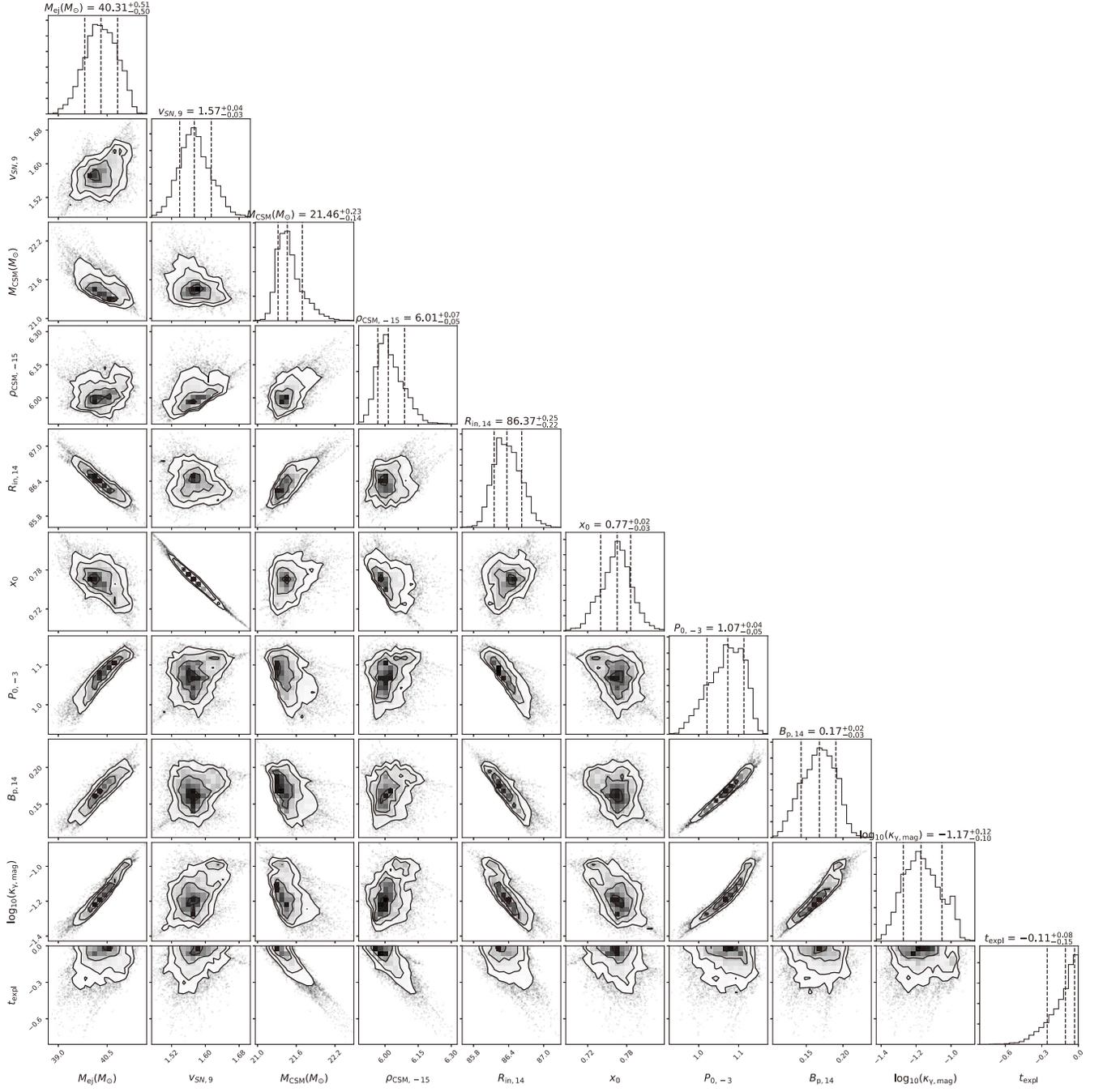}
\end{center}
\caption{The corner plot for the CSI plus magnetar model ($s=2$).}
\label{fig:CSI+mag2corner}
\end{figure}

The CSI plus magnetar model taking into account the gamma/X-ray leakage effect has 10 free parameters: $M_{\rm{ej}}$, $v_{\rm{SN}}$, $P_0$, $B_p$,
$\kappa_{\gamma,\rm{mag}}$, $M_{\rm{CSM}}$, $\rho_{\rm{CSM,in}}$, $R_{\rm CSM,in}$, $x_0$, and $t_{\rm{expl}}$. The LCs produced by the CSI plus magnetar
model are shown in solid blue curves in Figure \ref{fig:mag+CSI} and the best-fitting parameters are listed in Table \ref{tab:mag+CSI}.
Figure \ref{fig:CSI+mag2corner} is the corner plots showing the results of our MCMC parameter estimation for the CSI plus magnetar model ($s=2$).
The shape of the two dimensional projections of the posterior probability distributions indicates the correlations between the parameters: the closer to a circle,
the more independent the parameters are; the closer to a slender shape, the more correlated the parameters are. We note that there are many pairs of parameters show
degeneracies, which means there are many combinations of these degenerate parameters to achieve the same fitting result. This usually makes the best fitting parameters
more uncertain. Therefore, although the semi-analytic model successfully reproduce the luminosity evolution, one should treat these best fitting parameters with caution.

The results show that the dense-shell case cannot reproduce the plateau phase and the value of reduced $\chi^2$ ($\chi^2/\rm dof\sim5$) is higher
than that of the wind-like CSM model ($\chi^2/\rm dof\sim2$) which can reproduce the main peak, plateau, and late-time decline. The best-fitting parameters
are $M_{\rm ej}\sim 40~M_\odot$, $M_{\rm CSM}\sim 21~M_\odot$, and $v_{\rm SN}\sim 16,000\rm\ km\ s^{-1}$.
Using $E_{\rm SN}=\frac{(3-\delta)(n-3)}{2(5-\delta)(n-5)} M_{\rm ej}(x_0 v_{\rm SN})^2$, we derived the kinetic energy of the SN ejecta $
E_{\rm SN}\sim7.0\times10^{52}\rm\ erg$. Besides, a magnetar with initial spin period $P_0\sim1.1\rm\ ms$ and the
magnetic field strength $B_p\sim1.7\times10^{13}\rm\ G$ is needed to reproduce the late-time plateau and decline.

\subsection{The CSI Plus Fallback Model}

Here we propose an alternative model, in which the bolometric LC of ASASSN-15lh is supposed to be due to the jointing effect of the CSI and fallback
accretion. In this scenario, the main peak, the subsequent plateau, and the renewed decline at $\gtrsim200$ days since its detection are powered by CSI forward
shock, the CSI reverse shock, and the late-time black-hole fallback accretion, respectively. The LC of this energy source
model has the following form,
\begin{eqnarray}
  L(t) = \frac{1}{t_0} e^{-\frac{t}{t_0}} \int_{t_{\rm expl}}^{t+t_{\rm expl}} e^{\frac{t'}{t_0}} \epsilon [L_{\rm inp,FS}(t')+L_{\rm inp,RS}(t')] d t' + \frac{2}{t_d} e^{\frac{t^2}{t_d^2}} \int_{t_{\rm fb}-t_{\rm expl}}^{t+t_{\rm fb}-t_{\rm expl}} e^{\frac{t^2}{t_d^2}} L_{\rm inp,fb}(t') d t'.
\end{eqnarray}
where $t_0$ and $t_d$ are described in subsection \ref{fig:mag+CSI}, $t_{\rm{fb}}$ is the time when fallback accretion begins.
Generally, fallback accretion may happen at the beginning of explosion, when the materials with expansion velocity less than the escape velocity are eventually
accreted onto the central compact remnant. The accretion rate is usually flat at early times ($\lesssim10^3$ s), which is related to free-fall accretion,
proportional to $t^{-5/3}$ at late times \citep{Michel88,Chevalier89,Zhang08,Dexter13}. Numerical fallback simulations
\citep{Chevalier89,Zhang08,Dexter13} showed that the reverse shock forms when the ejecta meets the outer shell. The reverse shock can decelerate the ejecta and
enhance the fallback rate, but the late-time accretion rate is still proportional to $t^{-5/3}$.
Here, we consider a late-time enhanced fallback accretion due to the CSI reverse
shock and suppose that the enhanced fallback accretion occurs at $t_{\rm{fb}}$ after the SN explosion.

The accretion usually accompanied by an outflow carrying huge amount of energy.
The power input by the outflow associated with fallback accretion can be expressed by
\begin{equation}
  L_{\rm{inp,fb}}(t)=L_{\rm{fb,0}}\left(\frac{t}{t_{\rm{fb}}-t_{\rm expl}}\right)^{-5/3},
\end{equation}
where $L_{\rm{fb,0}}$ is the initial input power driven by fallback. Here, we assume the energy input from fallback is 100\% thermalized.

\begin{figure}[tbph]
\begin{center}
\includegraphics[width=0.45\textwidth,angle=0]{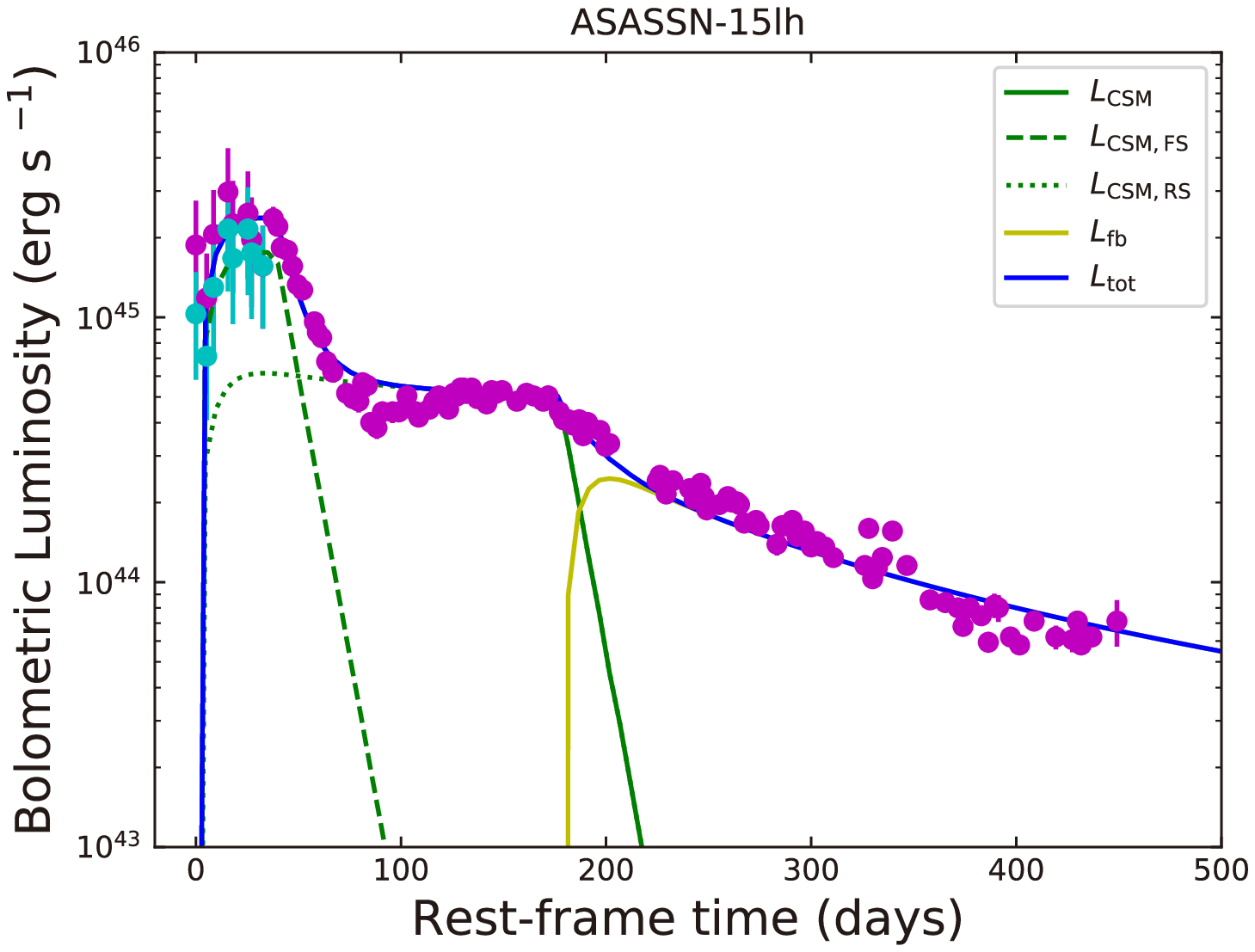}
\includegraphics[width=0.45\textwidth,angle=0]{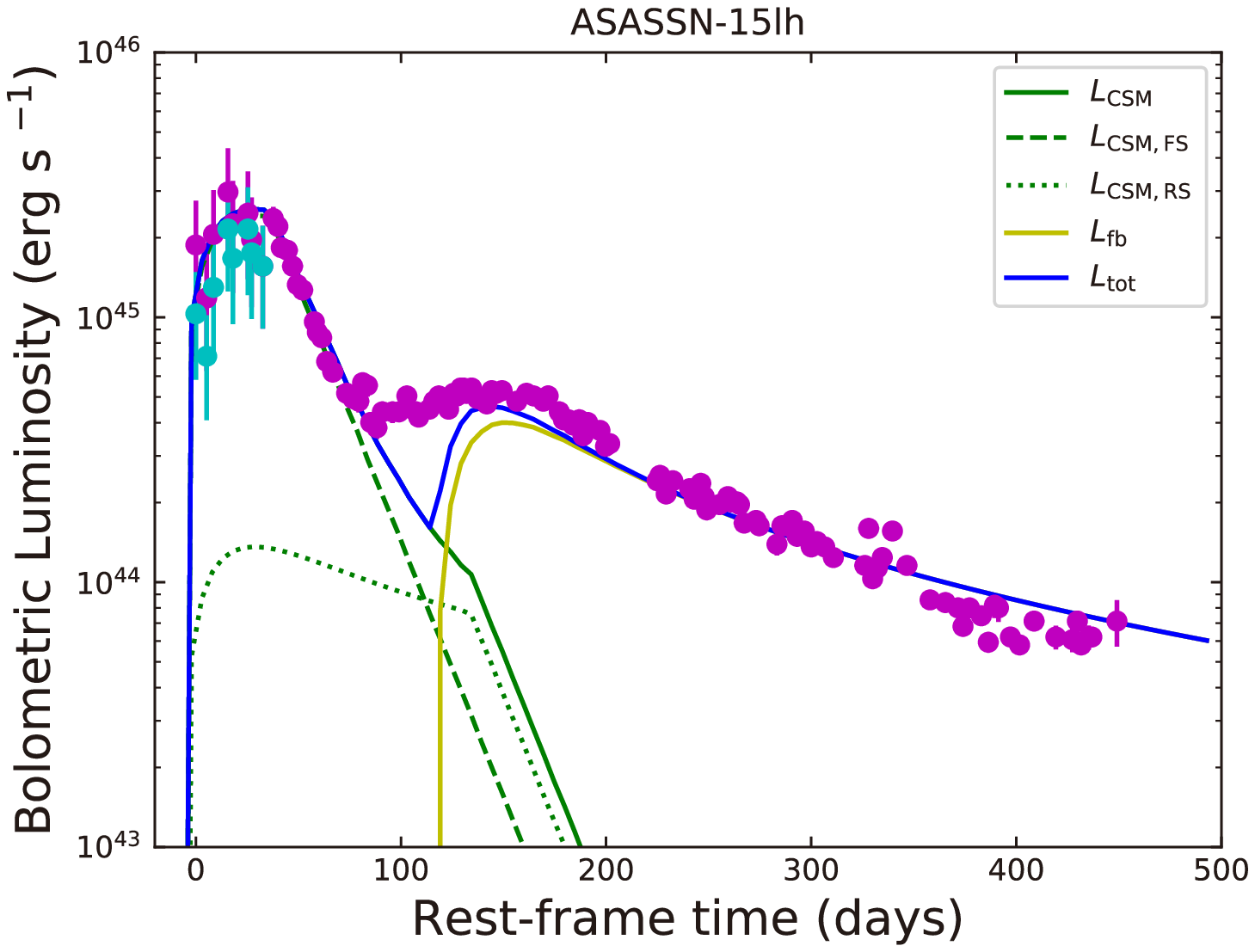}
\end{center}
\caption{The bolometric LCs of ASASSN-15lh reproduced by the CSI plus fallback model: $s = 0$ (the left panel) and $s = 2$ (the right panel). The total luminosity ($L_{\rm tot}$) are shown by solid blue curves. The contributions from the forward shock ($L_{\rm CSI,FS}$), the reverse shock ($L_{\rm CSI,RS}$), and the fallback accretion ($L_{\rm fb}$) are shown by dashed green, dotted green, and solid yellow curves, respectively. Data are taken from \cite{GodoyRivera17}.}
\label{fig:mag+CSI+fb}
\end{figure}

\begin{figure}[tbph]
\begin{center}
\includegraphics[width=1.0\textwidth,angle=0]{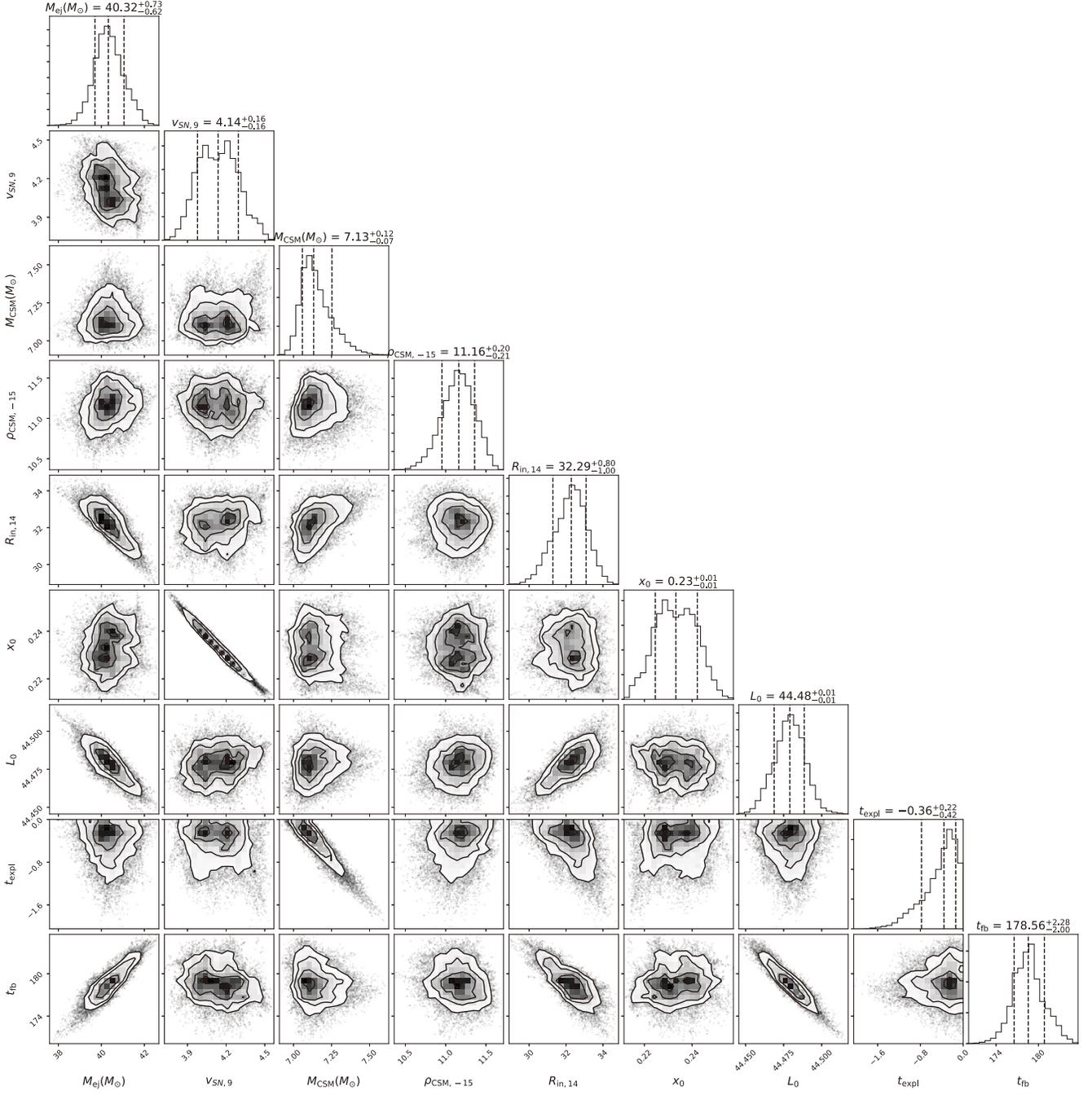}
\end{center}
\caption{The corner plot for the CSI plus fallback model ($s=0$).}
\label{fig:CSI+fb0corner}
\end{figure}

The CSI plus fallback model includes the following 9 parameters: $M_{\rm{ej}}$, $v_{\rm{SN}}$, $M_{\rm{CSM}}$, $\rho_{\rm{CSM,in}}$, $R_{\rm CSM,in}$,
$x_0$, $L_{\rm{fb,0}}$, $t_{\rm{fb}}$, and $t_{\rm expl}$.
The LCs produced by the model are shown in Figure \ref{fig:mag+CSI+fb} and the best-fitting parameters are listed in Table \ref{tab:mag+CSI+fb}.
Figure \ref{fig:CSI+fb0corner} is the corner plots showing the results of our Markov Chain Monte Carlo parameter estimation for the CSI plus fallback model ($s=0$).

The bolometric LC of ASASSN-15lh can only be fitted in the shell case with $\chi^2/\rm dof\sim4$. The wind case fails to reproduce the plateau and
the value of $\chi^2/\rm dof$ is $\sim9$. The derived values of $M_{\rm ej}$, $M_{\rm CSM}$, $v_{\rm SN}$,
and $E_{\rm SN}$ are $\sim 40~M_\odot$, $\sim 7~M_\odot$, $\sim 41,000\rm\ km\ s^{-1}$,
and $\sim4.5\times10^{52}\rm\ erg$, respectively.

Using $L_{\rm{fb,0}}$ and $t_{\rm{fb}}$, we can derive the total input energy
\begin{equation}
E_{\rm{fb}}=\int_{t_{\rm{fb}}-t_{\rm{expl}}}^{+\infty} L_{\rm{fb,0}}\left(\frac{t}{t_{\rm{fb}}-t_{\rm{expl}}}\right)^{-5/3}\ dt=1.5L_{\rm{fb,0}}(t_{\rm{fb}}-t_{\rm{expl}}).
\end{equation}
The value $E_{\rm{fb}}$ is $\sim7.0\times10^{51}$ erg. \cite{Dexter13} estimated that the typical value of the conversion efficiency of the
fallback accretion energy to outflow energy $\eta$ is $\sim10^{-3}$. Adopting this value, we find that the total mass of accretion $M_{\rm{fb}}=E_{\rm{fb}}/\eta c^2$ is
$\sim3.9~M_\odot$, indicating that only about a tenth of the ejecta finally fell back. The accretion rate range from $5.3~M_\odot\ \rm{yr}^{-1}$ to
$1.1~M_\odot\ \rm{yr}^{-1}$ after 180 to 450 days since the explosion.

\section{Discussion}
\label{sec:dis}

\subsection{Progenitor and Explosion Scenarios}

In Section \ref{sec:modeling}, we have demonstrated that both the ejecta-wind interaction plus magnetar spin-down with hard emission leakage model and the ejecta-shell interaction plus black hole fallback accretion model can reproduce the bolometric LC of ASASSN-15lh. The required total (SN ejecta + CSM) mass are $\sim 61 ~M_\odot$ and $\sim 47 ~M_\odot$ in the two models, respectively. The results are consistent with those of \cite{Chatzopoulos16}, suggesting that the explosion scenario would be a rapidly rotating pulsational pair-instability supernova (PPISN; \citealt{Chatzopoulos12}) or a progenitor related with a luminous blue variable (LBV; \citealt{Smith06}). For the CSI plus magnetar model, what one should caution is that the formation a magnetar rather than a black hole is difficult for a progenitor with mass large than $\sim 61 ~M_\odot$ \citep{Heger02}. It is believed that very massive stars would produce stellar-mass black holes. However, some magnetars might come from very massive progenitors (e.g., \citealt{Muno06} showed that an X-ray pulsar in the young massive cluster Westerlund 1 has a very massive progenitor $> 40~M_\odot$),  indicating that the CSI plus magnetar model cannot be ruled out by the large derived total mass.

After pulsational eruptions of hydrogen-poor materials, the progenitor might experiences energetic core-collapse explosion. The values of the kinetic energy inferred by the CSI plus magnetar model and the CSI plus fallback model are $\sim7.0\times10^{52}\rm\ erg$ and $\sim4.5\times10^{52}\rm\ erg$, respectively. These values are greater than the explosion energy $\sim10^{51}$ erg gotten by the neutrino-driven process. However, some so-called ``hypernovae" having kinetic energy comparable to the values inferred above have been discovered.\footnote{For instance, the kinitic energy of SN 1998bw, SN 2003dh, and SN 2003lw are $5\pm0.5\times10^{52}$ erg, $4\pm1\times10^{52}$ erg, and $6\pm1\times10^{52}$ erg, respectively (see Table 9 of \citealt{Hjorth12} and references therein).}

The large kinetic energy derived by the CSI plus magnetar model deserves further discussion. In the magnetar model, magnetar spinning-down  process  would convert a fraction of the rotational energy to the kinetic energy of the ejecta (see, e.g., \citealt{Wanglj16}). However, the fraction of the rotational energy converted to the kinetic energy depends on the spin-down timescale: a magnetar having a short spin-down timescale converts most of its rotational energy to the  kinetic energy of the ejecta, while a magnetar with long spin-down timescale converts a minor fraction of its rotational energy to the  kinetic energy of the ejecta and most of the rotational energy would be converted to radiation. The spin-down timescale of the magnetar in our model is $\sim216$ days, which is much larger than the diffusion timescale  $t_d$ which is $\sim100$ days. Therefore, the magnetar in our model would convert most of its rotational energy to radiation rather than the kinetic energy of the ejecta, i.e., the kinetic energy of the ejecta didn't come from the magnetar.

\subsection{The Mass Loss History of the Progenitor of ASASSN-15lh}

It is interesting to infer the mass-loss history of the progenitor of ASASSN-15lh. For the ejecta-wind interaction plus magnetar model,
assuming that the velocity of the wind ($v_{\rm w}$) is $100-1000\rm\ km\ s^{-1}$, we find that the mass loss rate $\dot M = 4 \pi v_{\rm w} q$
($q = \rho_{\rm CSM,in} R_{\rm CSM,in}^{2}$) of the stellar wind is $\dot M \sim 0.9-9~M_\odot\ \rm{yr}^{-1}$. This mass loss rate
is larger than the typical value $\sim 0.1-2~M_\odot\ \rm{yr}^{-1}$ of the other SN progenitors
\citep{Chugai03,Chugai04,Ofek14,Nyholm17}. But it can be explained in the light of ``superwinds" \citep{Moriya20}.

We can also estimate the time interval between the last eruption and the SN explosion is $\Delta t\sim2-25$ years by using $\Delta t=R_{\rm CSM,in}/v_{\rm w}$,
which is similar to those of SLSN PS1-12cil and SN 2012aa \citep{Li20}. Before this time, the progenitor has spent $M_{\rm CSM}/\dot M\sim 2.4-24$
years for stellar-wind mass loss.
For the ejecta-shell interaction plus fallback model, assuming that the velocity of the wind ($v_{\rm shell}$) is
$100-1000\rm\ km\ s^{-1}$, we see that the time interval between the eruption of shell and the SN explosion is $R_{\rm CSM,in}/v_{\rm shell}\sim1-10$ years.

\section{Conclusions}
\label{sec:con}

ASASSN-15lh might be the most luminous SN discovered to date. Its extremely high peak luminosity, long duration, and
the exotic shape of bolometric LC challenge the existing energy-source models for SLSNe. There are many studies for
the energy source of ASASSN-15lh, but the entire luminosity evolution of ASASSN-15lh has not been studied by the models
assuming that ASASSN-15lh is an SLSN.

In this paper, we fitted the whole bolometric LC of ASASSN-15lh by using several models taking into account the CSI contribution.
According to the physical properties of the CSM, the models are considered in two cases: the dense-shell CSM ($s=0$)
and wind-like CSM ($s=2$). The MCMC method was adopted to obtain the best fitting result.

We find that both the CSI model and the CSI plus magnetar model with full full gamma-ray/X-rays trapping
which were used by \cite{Chatzopoulos16} cannot reproduce the whole LC. To eliminate the late-time excess, we proposed
a CSI plus magnetar model with considering the leakage effect of gamma-ray/X-rays from the magnetar and found that the
model can well reproduce the overall bolometric LC of ASASSN-15lh if the CSM is a wind ($s=2$). The parameters
we derived are $M_{\rm ej}\sim 40~M_\odot$, $M_{\rm CSM}\sim 21~M_\odot$, $v_{\rm SN}\sim 16,000\rm\ km\ s^{-1}$, $P_0\sim1.1\rm\ ms$,
$B_p\sim1.7\times10^{13}\rm\ G$, $E_{\rm SN}\sim7.0\times10^{52}\rm\ erg$.
The stellar wind with mass-loss rate $\dot M \sim 0.9-9~M_\odot\ \rm{yr}^{-1}$ could have been expelled from progenitor in $\sim 2.4-24$ years,
and ceased at $\sim2-25$ years before explosion.

We also proposed a new hybrid model, the CSI plus fallback model, and applied it to the LC of ASASSN-15lh.
In this scenario, the luminosity evolution of ASASSN-15lh can only be explained in the shell case ($s=0$). The parameters we
derived are $M_{\rm ej}\sim 40~M_\odot$, $M_{\rm CSM}\sim 7~M_\odot$, $v_{\rm SN}\sim 41,000\rm\ km\ s^{-1}$,
and $E_{\rm SN}\sim4.5\times10^{52}\rm\ erg$. Assuming that the the conversion efficiency ($\eta$) of the fallback accretion to the outflow
is $\sim10^{-3}$ which is a typical value, we find that the total mass of accretion is $\sim3.9~M_\odot$.
Furthermore, the time interval between the eruption of shell and the SN explosion is $R_{\rm CSM,in}/v_w\sim1-10$ years.

These two models need a pre-SN eruption which might be powered by a PPISN eruption or
another mechanism. The CSI plus magnetar model favors a rapidly rotating progenitor that would left behind a millisecond
magnetar. To yield the huge amount of kinetic energy budget, both the CSI plus magnetar model and the CSI
plus fallback model need an extremely energetic explosion assembling that powering the explosions of hypernovae.

\acknowledgments
This work was supported by the National Key Research and Development Program of China (grant No. 2017YFA0402600) and the National Natural
Science Foundation of China (grant No. 11833003). S.Q.W. was supported by National Natural Science Foundation of China (Grant Nos. 11963001
and 11533003) and Guangxi Science Foundation (Grant No 2018GXNSFGA281007).

\clearpage

\begin{table*}[tbp]
\caption{Free Parameters and Priors in Three Models.}
\label{tab:prior}
\begin{center}
{\begin{tabular}{lccccccccccccccc}
\hline\hline
	&	\multicolumn{3}{c}{CSI model}	&	\multicolumn{3}{c}{CSI + magnetar model}	&	\multicolumn{3}{c}{CSI + fallback model}	\\
\hline
Parameter	&	Prior	&	Min	&	Max	&	Prior	&	Min	&	Max	&	Prior	&	Min	&	Max	\\
\hline
$M_{\mathrm{ej}}$ (M$_{\odot}$)	&	flat	&	0.1	&	50	&	flat	&	0.1	&	50	&	flat	&	0.1	&	50	\\
$v_{\mathrm{SN}}$ ($10^{9}$cm s$^{-1}$)	&	flat	&	1	&	5	&	flat	&	1	&	5	&	flat	&	1	&	5	\\
$M_{\mathrm{CSM}}$  (M$_{\odot}$) 	&	flat	&	0.1	&	50	&	flat	&	0.1	&	50	&	flat	&	0.1	&	50	\\
 $\rho_{\mathrm{CSM,in}}$ ($10^{-15}$g cm$^{-3}$)	&	flat	&	0.01	&	100	&	flat	&	0.01	&	100	&	flat	&	0.01	&	100	 \\
$R_{\mathrm{CSM,in}}$ ($10^{14}$cm)	&	flat	&	0.01	&	100	&	flat	&	0.01	&	100	&	flat	&	0.01	&	100	\\
$x_{\mathrm{0}}$	&	flat	&	0.01	&	1	&	flat	&	0.01	&	1	&	flat	&	0.01	&	1	\\
$P_0$ (ms)	&	-	&	-	&	-	&	flat	&	0.1	&	100	&	-	&	-	&	-	\\
$B_p$ ($10^{14}$~G)	&	-	&	-	&	-	&	flat	&	0.1	&	100	&	-	&	-	&	-	\\
$\kappa_{\gamma,\mathrm{mag}}$ (cm$^2$ g$^{-1}$)	&	-	&	-	&	-	&	log-flat	&	0.001	&	100	&	-	&	-	&	-	\\
$L_\mathrm{fb,0}$ ($10^{44}$erg s$^{-1}$)	&	-	&	-	&	-	&	-	&	-	&	-	&	log-flat	&	0.001	&	1000	\\
$t_\mathrm{expl}$ (days)	&	flat	&	$-$30	&	0	&	flat	&	$-$30	&	0	&	flat	&	$-$30	&	0	\\
$t_\mathrm{fb}$ (days)	&	-	&	-	&	-	&	-	&	-	&	-	&	flat	&	100	&	300	\\
\hline\hline										
\end{tabular}}
\end{center}
\end{table*}

\begin{table*}[tbp]
\tabcolsep=2pt
\caption{Parameters of the CSI plus magnetar model. The uncertainties are 1$\sigma$.}
\label{tab:mag+CSI}
\begin{center}
{\begin{tabular}{cccccccccccccccc}
\hline\hline																		
$s$	&	$M_{\mathrm{ej}}$	&	$v_{\mathrm{SN}}$	&	$M_{\mathrm{CSM}}$ 	&	 $\rho_{\mathrm{CSM,in}}$	&	$R_{\mathrm{CSM,in}}$	&	 $x_{\mathrm{0}}$	&	 $P_0$	&	$B_p$	&	 $\kappa_{\gamma,\mathrm{mag}}$	&	 $t_\mathrm{expl}$$^a$	&	$\chi^2/\mathrm{dof}$ \\
	&	(M$_{\odot}$)	&	($10^{9}$cm s$^{-1}$)	&	(M$_{\odot}$) 	&	 ($10^{-15}$g cm$^{-3}$)	&	($10^{14}$cm)	&		&	(ms)	&	 ($10^{14}$~G)	&	(cm$^2$ g$^{-1}$)	&	 (days)	&	\\
\hline\hline																						
0	&	$4.83^{+0.07}_{-0.10}$	&	$3.39^{+0.16}_{-0.12}$	&	$12.50^{+1.05}_{-0.22}$	&	$16.08^{+0.09}_{-0.32}$	&	$68.25^{+0.14}_{-0.85}$	&	 $0.43^{+0.01}_{-0.02}$	&	 $0.90^{+0.04}_{-0.02}$	&	$0.11^{+0.01}_{-0.01}$	&	 $0.87^{+0.10}_{-0.09}$	&	$-15.05^{+0.77}_{-0.23}$	&	 538.68/107\\
2	&	$40.31^{+0.51}_{-0.50}$	&	$1.57^{+0.04}_{-0.03}$	&	$21.46^{+0.23}_{-0.14}$	&	$6.01^{+0.07}_{-0.05}$	&	$86.37^{+0.25}_{-0.22}$	&	 $0.77^{+0.02}_{-0.03}$	&	 $1.07^{+0.04}_{-0.05}$	&	$0.17^{+0.02}_{-0.03}$	&	 $0.07^{+0.02}_{-0.01}$	&	$-0.11^{+0.08}_{-0.15}$	&	 203.06/107\\
\hline
\multicolumn{12}{l}{(Parameters from \cite{Chatzopoulos16})$^b$}\\
0	&	6.00	&	5.95	&	22.00	&	2.79	&	90.0	&	-	&	1.00	&	0.12	&	$\infty$	&	-	&	-\\
2	&	33.00	&	3.36	&	19.00	&	11.16	&	60.0	&	-	&	1.00	&	0.11	&	$\infty$	&	-	&	-\\
\hline\hline																		
\end{tabular}}
\end{center}
{$^a$The values of $t_\mathrm{expl}$ are with respect to the time of the first photometric observation.\\
$^b$The values of $x_{\mathrm{0}}$, $t_\mathrm{expl}$, and $\chi^2/\mathrm{dof}$ haven't been listed in \cite{Chatzopoulos16}. These authors did not take into account the leakage effect of hard emission, so $\kappa_{\gamma,\mathrm{mag}}$ is infinity.}
\end{table*}

\begin{table*}[tbp]
\tabcolsep=2pt
\caption{Parameters of the CSI plus fallback model. The uncertainties are 1$\sigma$.}
\label{tab:mag+CSI+fb}
\begin{center}
{\begin{tabular}{cccccccccccccccc}
\hline\hline																		
$s$	&	$M_{\mathrm{ej}}$	&	$v_{\mathrm{SN}}$	&	$M_{\mathrm{CSM}}$ 	&	 $\rho_{\mathrm{CSM,in}}$	&	$R_{\mathrm{CSM,in}}$	&	 $x_{\mathrm{0}}$	&	 $L_\mathrm{fb,0}$	&	 $t_\mathrm{expl}$$^a$	&	 $t_\mathrm{fb}$$^a$	&	$\chi^2/\mathrm{dof}$\\
	&	(M$_{\odot}$)	&	($10^{9}$cm s$^{-1}$)	&	(M$_{\odot}$) 	&	 ($10^{-15}$g cm$^{-3}$)	&	($10^{14}$cm)	&		&	($10^{44}$erg s$^{-1}$)	&	 (days)	&	(days)	&	\\
\hline\hline																				
0	&	$40.32^{+0.73}_{-0.62}$	&	$4.14^{+0.16}_{-0.16}$	&	$7.13^{+0.12}_{-0.07}$	&	$11.16^{+0.20}_{-0.21}$	&	$32.29^{+0.80}_{-1.00}$	&	 $0.23^{+0.01}_{-0.01}$	&	 $3.01^{+0.07}_{-0.07}$	&	$-0.36^{+0.22}_{-0.42}$	&	 $178.56^{+2.28}_{-2.00}$	&	442.58/108\\
2	&	$38.95^{+5.19}_{-7.27}$	&	$3.27^{+0.36}_{-0.29}$	&	$20.97^{+0.58}_{-0.74}$	&	$35.06^{+2.24}_{-0.84}$	&	$39.91^{+0.72}_{-1.38}$	&	 $0.32^{+0.02}_{-0.03}$	&	 $6.09^{+0.32}_{-0.30}$	&	$-7.14^{+0.35}_{-0.30}$	&	 $116.47^{+2.40}_{-3.68}$	&	1005.66/108\\
\hline\hline																		
\end{tabular}}
\end{center}
{$^a$The values of $t_\mathrm{expl}$ and $t_\mathrm{fb}$ are with respect to the time of the first photometric observation.}
\end{table*}

\clearpage

\clearpage

\clearpage

\clearpage

\clearpage

\begin{thebibliography}{}
\expandafter\ifx\csname natexlab\endcsname\relax\def\natexlab#1{#1}\fi

\bibitem[{{Angus} {et~al.}(2019){Angus}, {Smith}, {Sullivan}, {Inserra},
  {Wiseman}, {D'Andrea}, {Thomas}, {Nichol}, {Galbany}, {Childress}, {Asorey},
  {Brown}, {Casas}, {Castander}, {Curtin}, {Frohmaier}, {Glazebrook}, {Gruen},
  {Gutierrez}, {Kessler}, {Kim}, {Lidman}, {Macaulay}, {Nugent}, {Pursiainen},
  {Sako}, {Soares-Santos}, {Thomas}, {Abbott}, {Avila}, {Bertin}, {Brooks},
  {Buckley-Geer}, {Burke}, {Carnero Rosell}, {Carretero}, {da Costa}, {De
  Vicente}, {Desai}, {Diehl}, {Doel}, {Eifler}, {Flaugher}, {Fosalba},
  {Frieman}, {Garc{\'\i}a-Bellido}, {Gruendl}, {Gschwend}, {Hartley},
  {Hollowood}, {Honscheid}, {Hoyle}, {James}, {Kuehn}, {Kuropatkin}, {Lahav},
  {Lima}, {Maia}, {March}, {Marshall}, {Menanteau}, {Miller}, {Miquel}, {Ogand
  o}, {Plazas}, {Romer}, {Sanchez}, {Schindler}, {Schubnell}, {Sobreira},
  {Suchyta}, {Swanson}, {Tarle}, {Thomas}, {Tucker}, \& {DES
  Collaboration}}]{Angus19}
{Angus}, C.~R., {Smith}, M., {Sullivan}, M., {et~al.} 2019, \mnras, 487, 2215

\bibitem[{{Barkat} {et~al.}(1967){Barkat}, {Rakavy}, \& {Sack}}]{Barkat67}
{Barkat}, Z., {Rakavy}, G., \& {Sack}, N. 1967, \prl, 18, 379

\bibitem[{{Bersten} {et~al.}(2016){Bersten}, {Benvenuto}, {Orellana}, \&
  {Nomoto}}]{Bersten16}
{Bersten}, M.~C., {Benvenuto}, O.~G., {Orellana}, M., \& {Nomoto}, K. 2016,
  \apjl, 817, L8

\bibitem[{{Brown} {et~al.}(2013){Brown}, {Baliber}, {Bianco}, {Bowman},
  {Burleson}, {Conway}, {Crellin}, {Depagne}, {De Vera}, {Dilday}, {Dragomir},
  {Dubberley}, {Eastman}, {Elphick}, {Falarski}, {Foale}, {Ford}, {Fulton},
  {Garza}, {Gomez}, {Graham}, {Greene}, {Haldeman}, {Hawkins}, {Haworth},
  {Haynes}, {Hidas}, {Hjelstrom}, {Howell}, {Hygelund}, {Lister}, {Lobdill},
  {Martinez}, {Mullins}, {Norbury}, {Parrent}, {Paulson}, {Petry}, {Pickles},
  {Posner}, {Rosing}, {Ross}, {Sand}, {Saunders}, {Shobbrook}, {Shporer},
  {Street}, {Thomas}, {Tsapras}, {Tufts}, {Valenti}, {Vander Horst}, {Walker},
  {White}, \& {Willis}}]{Brown13}
{Brown}, T.~M., {Baliber}, N., {Bianco}, F.~B., {et~al.} 2013, \pasp, 125, 1031

\bibitem[{{Chatzopoulos} {et~al.}(2012){Chatzopoulos}, {Wheeler}, \&
  {Vinko}}]{Chatzopoulos12}
{Chatzopoulos}, E., {Wheeler}, J.~C., \& {Vinko}, J. 2012, \apj, 746, 121

\bibitem[{{Chatzopoulos} {et~al.}(2013){Chatzopoulos}, {Wheeler}, {Vinko},
  {Horvath}, \& {Nagy}}]{Chatzopoulos13}
{Chatzopoulos}, E., {Wheeler}, J.~C., {Vinko}, J., {Horvath}, Z.~L., \& {Nagy},
  A. 2013, \apj, 773, 76

\bibitem[{{Chatzopoulos} {et~al.}(2016){Chatzopoulos}, {Wheeler}, {Vinko},
  {Nagy}, {Wiggins}, \& {Even}}]{Chatzopoulos16}
{Chatzopoulos}, E., {Wheeler}, J.~C., {Vinko}, J., {et~al.} 2016, \apj, 828, 94

\bibitem[{{Chevalier}(1982)}]{Chevalier82}
{Chevalier}, R.~A. 1982, \apj, 258, 790

\bibitem[{{Chevalier}(1989)}]{Chevalier89}
---. 1989, \apj, 346, 847

\bibitem[{{Chevalier} \& {Fransson}(1994)}]{Chevalier94}
{Chevalier}, R.~A., \& {Fransson}, C. 1994, \apj, 420, 268

\bibitem[{{Chevalier} \& {Irwin}(2011)}]{Chevalier11}
{Chevalier}, R.~A., \& {Irwin}, C.~M. 2011, \apjl, 729, L6

\bibitem[{{Chomiuk} {et~al.}(2011){Chomiuk}, {Chornock}, {Soderberg}, {Berger},
  {Chevalier}, {Foley}, {Huber}, {Narayan}, {Rest}, {Gezari}, {Kirshner},
  {Riess}, {Rodney}, {Smartt}, {Stubbs}, {Tonry}, {Wood-Vasey}, {Burgett},
  {Chambers}, {Czekala}, {Flewelling}, {Forster}, {Kaiser}, {Kudritzki},
  {Magnier}, {Martin}, {Morgan}, {Neill}, {Price}, {Roth}, {Sanders}, \&
  {Wainscoat}}]{Chomiuk11}
{Chomiuk}, L., {Chornock}, R., {Soderberg}, A.~M., {et~al.} 2011, \apj, 743,
  114

\bibitem[{{Chugai} \& {Danziger}(2003)}]{Chugai03}
{Chugai}, N.~N., \& {Danziger}, I.~J. 2003, Astronomy Letters, 29, 649

\bibitem[{{Chugai} {et~al.}(2004){Chugai}, {Blinnikov}, {Cumming}, {Lundqvist},
  {Bragaglia}, {Filippenko}, {Leonard}, {Matheson}, \& {Sollerman}}]{Chugai04}
{Chugai}, N.~N., {Blinnikov}, S.~I., {Cumming}, R.~J., {et~al.} 2004, \mnras,
  352, 1213

\bibitem[{{Dai} {et~al.}(2016){Dai}, {Wang}, {Wang}, {Wang}, \& {Yu}}]{Dai16}
{Dai}, Z.~G., {Wang}, S.~Q., {Wang}, J.~S., {Wang}, L.~J., \& {Yu}, Y.~W. 2016,
  \apj, 817, 132

\bibitem[{{De Cia} {et~al.}(2018){De Cia}, {Gal-Yam}, {Rubin}, {Leloudas},
  {Vreeswijk}, {Perley}, {Quimby}, {Yan}, {Sullivan}, {Fl{\"o}rs}, {Sollerman},
  {Bersier}, {Cenko}, {Gal-Yam}, {Maguire}, {Ofek}, {Prentice}, {Schulze},
  {Spyromilio}, {Valenti}, {Arcavi}, {Corsi}, {Howell}, {Mazzali}, {Kasliwal},
  {Taddia}, \& {Yaron}}]{DeCia18}
{De Cia}, A., {Gal-Yam}, A., {Rubin}, A., {et~al.} 2018, \apj, 860, 100

\bibitem[{{Dexter} \& {Kasen}(2013)}]{Dexter13}
{Dexter}, J., \& {Kasen}, D. 2013, \apj, 772, 30

\bibitem[{{Dong} {et~al.}(2016){Dong}, {Shappee}, {Prieto}, {Jha}, {Stanek},
  {Holoien}, {Kochanek}, {Thompson}, {Morrell}, {Thompson}, {Basu}, {Beacom},
  {Bersier}, {Brimacombe}, {Brown}, {Bufano}, {Chen}, {Conseil}, {Danilet},
  {Falco}, {Grupe}, {Kiyota}, {Masi}, {Nicholls}, {Olivares E.}, {Pignata},
  {Pojmanski}, {Simonian}, {Szczygiel}, \& {Wo{\'z}niak}}]{Dong16}
{Dong}, S., {Shappee}, B.~J., {Prieto}, J.~L., {et~al.} 2016, Science, 351, 257

\bibitem[{{Foreman-Mackey} {et~al.}(2013){Foreman-Mackey}, {Hogg}, {Lang}, \&
  {Goodman}}]{emcee}
{Foreman-Mackey}, D., {Hogg}, D.~W., {Lang}, D., \& {Goodman}, J. 2013, \pasp,
  125, 306

\bibitem[{{Gal-Yam}(2012)}]{GalYam12}
{Gal-Yam}, A. 2012, Science, 337, 927

\bibitem[{{Gal-Yam}(2019)}]{GalYam19}
---. 2019, \araa, 57, 305

\bibitem[{{Gehrels} {et~al.}(2004){Gehrels}, {Chincarini}, {Giommi}, {Mason},
  {Nousek}, {Wells}, {White}, {Barthelmy}, {Burrows}, {Cominsky}, {Hurley},
  {Marshall}, {M{\'e}sz{\'a}ros}, {Roming}, {Angelini}, {Barbier}, {Belloni},
  {Campana}, {Caraveo}, {Chester}, {Citterio}, {Cline}, {Cropper}, {Cummings},
  {Dean}, {Feigelson}, {Fenimore}, {Frail}, {Fruchter}, {Garmire}, {Gendreau},
  {Ghisellini}, {Greiner}, {Hill}, {Hunsberger}, {Krimm}, {Kulkarni}, {Kumar},
  {Lebrun}, {Lloyd-Ronning}, {Markwardt}, {Mattson}, {Mushotzky}, {Norris},
  {Osborne}, {Paczynski}, {Palmer}, {Park}, {Parsons}, {Paul}, {Rees},
  {Reynolds}, {Rhoads}, {Sasseen}, {Schaefer}, {Short}, {Smale}, {Smith},
  {Stella}, {Tagliaferri}, {Takahashi}, {Tashiro}, {Townsley}, {Tueller},
  {Turner}, {Vietri}, {Voges}, {Ward}, {Willingale}, {Zerbi}, \&
  {Zhang}}]{Gehrels04}
{Gehrels}, N., {Chincarini}, G., {Giommi}, P., {et~al.} 2004, \apj, 611, 1005

\bibitem[{{Godoy-Rivera} {et~al.}(2017){Godoy-Rivera}, {Stanek}, {Kochanek},
  {Chen}, {Dong}, {Prieto}, {Shappee}, {Jha}, {Foley}, {Pan}, {Holoien},
  {Thompson}, {Grupe}, \& {Beacom}}]{GodoyRivera17}
{Godoy-Rivera}, D., {Stanek}, K.~Z., {Kochanek}, C.~S., {et~al.} 2017, \mnras,
  466, 1428

\bibitem[{{Heger} {et~al.}(2003){Heger}, {Fryer}, {Woosley}, {Langer}, \&
  {Hartmann}}]{Heger03}
{Heger}, A., {Fryer}, C.~L., {Woosley}, S.~E., {Langer}, N., \& {Hartmann},
  D.~H. 2003, \apj, 591, 288

\bibitem[{{Heger} \& {Woosley}(2002)}]{Heger02}
{Heger}, A., \& {Woosley}, S.~E. 2002, \apj, 567, 532

\bibitem[{{Hjorth} \& {Bloom}(2012)}]{Hjorth12}
{Hjorth}, J., \& {Bloom}, J.~S. 2012, {The Gamma-Ray Burst - Supernova
  Connection}, 169--190

\bibitem[{{Inserra}(2019)}]{Inserra19}
{Inserra}, C. 2019, Nature Astronomy, 3, 697

\bibitem[{{Inserra} {et~al.}(2013){Inserra}, {Smartt}, {Jerkstrand}, {Valenti},
  {Fraser}, {Wright}, {Smith}, {Chen}, {Kotak}, \& {Pastorello}}]{Inserra13}
{Inserra}, C., {Smartt}, S.~J., {Jerkstrand}, A., {et~al.} 2013, \apj, 770, 128

\bibitem[{{Kasen} \& {Bildsten}(2010)}]{Kasen10}
{Kasen}, D., \& {Bildsten}, L. 2010, \apj, 717, 245

\bibitem[{{Kozyreva} {et~al.}(2016){Kozyreva}, {Hirschi}, {Blinnikov}, \& {den
  Hartogh}}]{Kozyreva16}
{Kozyreva}, A., {Hirschi}, R., {Blinnikov}, S., \& {den Hartogh}, J. 2016,
  \mnras, 459, L21

\bibitem[{{Kr{\"u}hler} {et~al.}(2018){Kr{\"u}hler}, {Fraser}, {Leloudas},
  {Schulze}, {Stone}, {van Velzen}, {Amorin}, {Hjorth}, {Jonker}, {Kann},
  {Kim}, {Kuncarayakti}, {Mehner}, \& {Nicuesa Guelbenzu}}]{Kruhler18}
{Kr{\"u}hler}, T., {Fraser}, M., {Leloudas}, G., {et~al.} 2018, \aap, 610, A14

\bibitem[{{Leloudas} {et~al.}(2016){Leloudas}, {Fraser}, {Stone}, {van Velzen},
  {Jonker}, {Arcavi}, {Fremling}, {Maund}, {Smartt}, {Kr{\`i}hler},
  {Miller-Jones}, {Vreeswijk}, {Gal-Yam}, {Mazzali}, {De Cia}, {Howell},
  {Inserra}, {Patat}, {de Ugarte Postigo}, {Yaron}, {Ashall}, {Bar},
  {Campbell}, {Chen}, {Childress}, {Elias-Rosa}, {Harmanen}, {Hosseinzadeh},
  {Johansson}, {Kangas}, {Kankare}, {Kim}, {Kuncarayakti}, {Lyman}, {Magee},
  {Maguire}, {Malesani}, {Mattila}, {McCully}, {Nicholl}, {Prentice},
  {Romero-Ca{\~n}izales}, {Schulze}, {Smith}, {Sollerman}, {Sullivan},
  {Tucker}, {Valenti}, {Wheeler}, \& {Young}}]{Leloudas16}
{Leloudas}, G., {Fraser}, M., {Stone}, N.~C., {et~al.} 2016, Nature Astronomy,
  1, 0002

\bibitem[{{Li} {et~al.}(2020){Li}, {Wang}, {Liu}, {Wang}, {Liang}, \&
  {Dai}}]{Li20}
{Li}, L., {Wang}, S.-Q., {Liu}, L.-D., {et~al.} 2020, arXiv e-prints,
  arXiv:2001.09463

\bibitem[{{Liu} {et~al.}(2018){Liu}, {Wang}, {Wang}, \& {Dai}}]{Liu18}
{Liu}, L.-D., {Wang}, L.-J., {Wang}, S.-Q., \& {Dai}, Z.-G. 2018, \apj, 856, 59

\bibitem[{{Lunnan} {et~al.}(2018){Lunnan}, {Chornock}, {Berger}, {Jones},
  {Rest}, {Czekala}, {Dittmann}, {Drout}, {Foley}, {Fong}, {Kirshner},
  {Laskar}, {Leibler}, {Margutti}, {Milisavljevic}, {Narayan}, {Pan}, {Riess},
  {Roth}, {Sanders}, {Scolnic}, {Smartt}, {Smith}, {Chambers}, {Draper},
  {Flewelling}, {Huber}, {Kaiser}, {Kudritzki}, {Magnier}, {Metcalfe},
  {Wainscoat}, {Waters}, \& {Willman}}]{Lunnan18}
{Lunnan}, R., {Chornock}, R., {Berger}, E., {et~al.} 2018, \apj, 852, 81

\bibitem[{{Margutti} {et~al.}(2017){Margutti}, {Metzger}, {Chornock},
  {Milisavljevic}, {Berger}, {Blanchard}, {Guidorzi}, {Migliori}, {Kamble},
  {Lunnan}, {Nicholl}, {Coppejans}, {Dall'Osso}, {Drout}, {Perna}, \&
  {Sbarufatti}}]{Margutti17}
{Margutti}, R., {Metzger}, B.~D., {Chornock}, R., {et~al.} 2017, \apj, 836, 25

\bibitem[{{Metzger} {et~al.}(2015){Metzger}, {Margalit}, {Kasen}, \&
  {Quataert}}]{Metzger15}
{Metzger}, B.~D., {Margalit}, B., {Kasen}, D., \& {Quataert}, E. 2015, \mnras,
  454, 3311

\bibitem[{{Michel}(1988)}]{Michel88}
{Michel}, F.~C. 1988, \nat, 333, 644

\bibitem[{{Moriya} {et~al.}(2020){Moriya}, {Mazzali}, \& {Pian}}]{Moriya20}
{Moriya}, T.~J., {Mazzali}, P.~A., \& {Pian}, E. 2020, \mnras, 491, 1384

\bibitem[{{Moriya} {et~al.}(2018){Moriya}, {Sorokina}, \&
  {Chevalier}}]{Moriya18}
{Moriya}, T.~J., {Sorokina}, E.~I., \& {Chevalier}, R.~A. 2018, \ssr, 214, 59

\bibitem[{{Mummery} \& {Balbus}(2020)}]{Mummery20}
{Mummery}, A., \& {Balbus}, S.~A. 2020, \mnras, 497, L13

\bibitem[{{Muno} {et~al.}(2006){Muno}, {Clark}, {Crowther}, {Dougherty}, {de
  Grijs}, {Law}, {McMillan}, {Morris}, {Negueruela}, {Pooley}, {Portegies
  Zwart}, \& {Yusef-Zadeh}}]{Muno06}
{Muno}, M.~P., {Clark}, J.~S., {Crowther}, P.~A., {et~al.} 2006, \apjl, 636,
  L41

\bibitem[{{Nicholl} {et~al.}(2013){Nicholl}, {Smartt}, {Jerkstrand}, {Inserra},
  {McCrum}, {Kotak}, {Fraser}, {Wright}, {Chen}, \& {Smith}}]{Nicholl13}
{Nicholl}, M., {Smartt}, S.~J., {Jerkstrand}, A., {et~al.} 2013, \nat, 502, 346

\bibitem[{{Nicholl} {et~al.}(2014){Nicholl}, {Smartt}, {Jerkstrand}, {Inserra},
  {Anderson}, {Baltay}, {Benetti}, {Chen}, {Elias-Rosa}, \&
  {Feindt}}]{Nicholl14}
---. 2014, \mnras, 444, 2096

\bibitem[{{Nyholm} {et~al.}(2017){Nyholm}, {Sollerman}, {Taddia}, {Fremling},
  {Moriya}, {Ofek}, {Gal-Yam}, {De Cia}, {Roy}, \& {Kasliwal}}]{Nyholm17}
{Nyholm}, A., {Sollerman}, J., {Taddia}, F., {et~al.} 2017, \aap, 605, A6

\bibitem[{{Ofek} {et~al.}(2014){Ofek}, {Sullivan}, {Shaviv}, {Steinbok},
  {Arcavi}, {Gal-Yam}, {Tal}, {Kulkarni}, {Nugent}, {Ben-Ami}, {Kasliwal},
  {Cenko}, {Laher}, {Surace}, {Bloom}, {Filippenko}, {Silverman}, \&
  {Yaron}}]{Ofek14}
{Ofek}, E.~O., {Sullivan}, M., {Shaviv}, N.~J., {et~al.} 2014, \apj, 789, 104

\bibitem[{{Quimby}(2014)}]{Quimby14}
{Quimby}, R.~M. 2014, in IAU Symposium, Vol. 296, Supernova Environmental
  Impacts, ed. A.~{Ray} \& R.~A. {McCray}, 68--76

\bibitem[{{Quimby} {et~al.}(2011){Quimby}, {Kulkarni}, {Kasliwal}, {Gal-Yam},
  {Arcavi}, {Sullivan}, {Nugent}, {Thomas}, {Howell}, {Nakar}, {Bildsten},
  {Theissen}, {Law}, {Dekany}, {Rahmer}, {Hale}, {Smith}, {Ofek}, {Zolkower},
  {Velur}, {Walters}, {Henning}, {Bui}, {McKenna}, {Poznanski}, {Cenko}, \&
  {Levitan}}]{Quimby11}
{Quimby}, R.~M., {Kulkarni}, S.~R., {Kasliwal}, M.~M., {et~al.} 2011, \nat,
  474, 487

\bibitem[{{Rakavy} \& {Shaviv}(1967)}]{Rakavy67}
{Rakavy}, G., \& {Shaviv}, G. 1967, \apj, 148, 803

\bibitem[{{Roming} {et~al.}(2005){Roming}, {Kennedy}, {Mason}, {Nousek}, {Ahr},
  {Bingham}, {Broos}, {Carter}, {Hancock}, {Huckle}, {Hunsberger}, {Kawakami},
  {Killough}, {Koch}, {McLelland}, {Smith}, {Smith}, {Soto}, {Boyd},
  {Breeveld}, {Holland}, {Ivanushkina}, {Pryzby}, {Still}, \&
  {Stock}}]{Roming05}
{Roming}, P. W.~A., {Kennedy}, T.~E., {Mason}, K.~O., {et~al.} 2005, \ssr, 120,
  95

\bibitem[{{Shappee} {et~al.}(2014){Shappee}, {Prieto}, {Grupe}, {Kochanek},
  {Stanek}, {De Rosa}, {Mathur}, {Zu}, {Peterson}, {Pogge}, {Komossa}, {Im},
  {Jencson}, {Holoien}, {Basu}, {Beacom}, {Szczygie{\l}}, {Brimacombe},
  {Adams}, {Campillay}, {Choi}, {Contreras}, {Dietrich}, {Dubberley},
  {Elphick}, {Foale}, {Giustini}, {Gonzalez}, {Hawkins}, {Howell}, {Hsiao},
  {Koss}, {Leighly}, {Morrell}, {Mudd}, {Mullins}, {Nugent}, {Parrent},
  {Phillips}, {Pojmanski}, {Rosing}, {Ross}, {Sand}, {Terndrup}, {Valenti},
  {Walker}, \& {Yoon}}]{Shappee14}
{Shappee}, B.~J., {Prieto}, J.~L., {Grupe}, D., {et~al.} 2014, \apj, 788, 48

\bibitem[{{Smith} \& {Owocki}(2006)}]{Smith06}
{Smith}, N., \& {Owocki}, S.~P. 2006, \apjl, 645, L45

\bibitem[{{Sukhbold} \& {Woosley}(2016)}]{Sukhbold16}
{Sukhbold}, T., \& {Woosley}, S.~E. 2016, \apjl, 820, L38

\bibitem[{{Wang} {et~al.}(2016){Wang}, {Wang}, {Dai}, {Xu}, {Han}, {Wu}, \&
  {Wei}}]{Wanglj16}
{Wang}, L.-J., {Wang}, S.~Q., {Dai}, Z.~G., {et~al.} 2016, \apj, 821, 22

\bibitem[{{Wang} {et~al.}(2019{\natexlab{a}}){Wang}, {Wang}, {Cano}, {Wang},
  {Liu}, {Dai}, {Deng}, {Yu}, {Li}, {Song}, {Qiu}, \& {Wei}}]{Wanglj19}
{Wang}, L.~J., {Wang}, X.~F., {Cano}, Z., {et~al.} 2019{\natexlab{a}}, \mnras,
  489, 1110

\bibitem[{{Wang} {et~al.}(2019{\natexlab{b}}){Wang}, {Wang}, \& {Dai}}]{Wang19}
{Wang}, S.-Q., {Wang}, L.-J., \& {Dai}, Z.-G. 2019{\natexlab{b}}, Research in
  Astronomy and Astrophysics, 19, 063

\bibitem[{{Wang} {et~al.}(2015){Wang}, {Wang}, {Dai}, \& {Wu}}]{Wang15}
{Wang}, S.~Q., {Wang}, L.~J., {Dai}, Z.~G., \& {Wu}, X.~F. 2015, \apj, 799, 107

\bibitem[{{Zhang} {et~al.}(2008){Zhang}, {Woosley}, \& {Heger}}]{Zhang08}
{Zhang}, W., {Woosley}, S.~E., \& {Heger}, A. 2008, \apj, 679, 639

\end{thebibliography}
\end{document}